\documentstyle[12pt]{article}
\newcommand{\bq}{\begin{equation}}
\newcommand{\eq}{\end{equation}}
\def\fr{\frac}
\def\be{\begin{equation}}
\def\ee{\end{equation}}
\begin{document}
\parskip=5pt
\baselineskip=16pt
\bigskip
\bigskip
\bigskip
\medskip

{\raggedleft {\makebox[2.7cm][1]{\it November 1994}}\\}

\ \\ \\ \\ \\
\begin{center}
{\Large\bf  On inserter Regularization  Method{\footnote{Talk by H.Y.Guo at ITP
workshop on Gauge Theory and WZW model during Dec. 5-10, also a concise version
of hep-th/9412034.}}}
\end{center}

\vspace{10ex}
\begin{center}

{\large Han-Ying Guo}\\
\vskip 3mm
Max-Planck-Institut f\"ur Mathematik, D-53225 Bonn, Germany;\\
\vskip 2mm
and\\
\vskip 2mm
Institute of Theoretical Physics, Academia Sinica,
P.O. Box 2735, Beijing 100080, China.
\footnote {Permanent address.}\\

\vskip 6mm
{\large Yu Cai ~and~ Hong-Bo Teng{\footnote{Email: hyguo@itp.ac.cn,
caiyu@itp.ac.cn and tenghb@itp.ac.cn for HYG, YC, and HBT respectively.}}}\\
\vskip 2mm
Institute of Theoretical Physics, Academia Sinica,
P.O. Box 2735, Beijing 100080, China.\\

\end{center}
\vspace{8ex}
\vspace{8ex}

\centerline{\it ABSTRACT}
\vskip 1mm
\begin{center}
\begin{minipage}{142mm}
{\it There exist certain intrinsic relations between the ultraviolet divergent
graphs and the convergent ones at the same loop order in  renormalizable
quantum field theories. Whereupon we present a new method, the inserter
regularization method, to regulate
 those divergent graphs.  In this letter, we demonstrate this method with
 the  $\phi^4$ theory and QED at the
one loop order.  Some applications to SUSY-models are also made at the one loop
order, which shows that supersymmetry is preserved manifestly and
consistently.}

\end{minipage}
\end{center}
\newpage

As is well known, various regularization  schemes
have been developed in the quantum field theory. However, the topic is
still an important and fundamental issues under investigation.
One of the most challenging problems is perhaps how to preserve  all
 symmetries and topological properties manifestly and consistently.

 It has been found that there exist certain intrinsic relations between
the  ultraviolet  divergent
graphs and the convergent ones at the same loop order in  renormalizable
QFT [1-5]. Whereupon we should be able to establish a new method, the intrinsic
regularization method,  for regularization
of those divergent graphs.
The key point of the new method is
based upon the following simple observation:
For a given  ultraviolet   divergent function at certain loop order in a
 renormalizable QFT, there always exists
a set of convergent functions at the same loop order such that their Feynman
graphs share
same loop skeleton  and the main difference
is that the convergent ones have additional vertices of certain kinds and the
original
one is the case without these vertices. This is an intrinsic
relation between the original  ultraviolet  divergent graph and the convergent
ones in the
QFT. It is this relation that indicates it is possible to introduce the
regulated function for the divergent one with the help of those convergent
 ones so that the potentially divergent
integral of the graph can be rendered  finite while for the limiting case of
the
number of the additional vertices $q\to 0$
 the divergence again becomes manifest in pole(s) of $q$.

It is very simple why there always exists such kind of intrinsic relations in
renormalizable QFTs. Let us consider some Feynman graphs at the
$L$ loop order with  $I$ internal lines of any kind and  $V$ vertices of any
 kind. The Euler formula
$$ L-I+V = 1$$
shows that for fixed $L$ $I$ increases the same as $V$ does so that the
superficial degree of divergence decreases. Therefore, for a given divergent
Feynman graph at certain loop order, the Euler formula insures that one
can always reach a set of convergent graphs in a suitable  perturbation
 expansion series in the order of some coupling constant
, which always appears with a vertex of certain kind, as long as the original
divergent graph is included in the series. In fact, the Euler formula is
a cornerstone of the intrinsic regularization method.
In general, however,
the procedure may be very involved. The aim of the inserter method to be
presented in this letter is just to simplify the procedure.

To be concrete, let us consider a 1PI graph with $I$ internal lines at one loop
order in
the $\phi^4$ theory. Its superficial degree of divergences in the momentum
space is
$$
\delta=4-2I.
$$
When $I=1$ or $2$, the graph is divergent. Obviously, there exists such kind of
graphs that they have additional $q$ four-$\phi$-vertices in the internal
lines. Then  the
divergent degree of the new 1PI graphs becomes
$$
\delta'=4-2(I+q).
$$
If $q$ is large enough, the new graphs are convergent and the original
divergent one is the case of $q=0$.  Thus,  an intrinsic relation has been
reached between the  original divergent 1PI graph and the
new convergent ones at the same loop order.

In the inserter regularization method,  we take all external
lines in the additional vertices with  zero momenta and call such a vertex
an {\it inserter}. Thus those convergent graphs can simply be regarded as the
ones given by suitably inserting $q$-inserters in all internal lines in the
given divergent graph and
 the powers of the propagators are simply raised so as to those integrals for
 the new  graphs become convergent.
It is clear that these new graphs share the same loop skeleton with the
original divergent one and the main differences from the original one are the
number of inserters as well as the dimension in mass and the order in the
 coupling constant due to the insertions. Thus
it is possible to regulate the
divergent graph based upon this intrinsic relation as long as
 we may get rid of those differences and deal with those convergent
 functions on an equal footing. To this end we introduce a
  well-defined convergent 1PI function, the
 regulated function, by taking the
arithmetical average of those convergent 1PI functions and changing their
 dimension in mass, their order in $\lambda$ etc. to that in
the original divergent function.
 Thus this new function renders the divergent
integral finite. Evaluating it and continuing $q$
analytically from the integer to the complex number, the  divergent function
of the original 1PI
 graph is recovered as
the $q \to 0$
limiting case of such a regulated function.

Therefore the main steps of the method for the $\phi^4$ theory
are as
follows.
First, an  inserter should be constructed. As mentioned above, it is a
four-$\phi$-vertex with two zero momentum external lines whose Feynman rule is
the same as the vertex
\be
I^{\phi}(p)=-i\lambda.
\ee
For a given  1PI $n$-point
divergent function at the one loop order $\Gamma^{(n)}(p_1,~\cdots,~p_n)$,
 we consider all $n+2q$-point functions
$\Gamma^{(n+2q)}(p_1,~\cdots,~p_{n}; ~q)$ which are  the
amplitudes
of the graphs corresponding to all possible $q$ insertions of the inserter in
the internal lines
 of the given $n$-point graph.  If
$q$ is large enough, $\Gamma^{(n+2q)}(p_1,~\cdots,~p_{n}; ~q)$ become
convergent. And for $q=0$ it is the case of original
$n$-point function.   With the help of this relation, we introduce a new
function by taking the arithmetical average of these convergent functions, i.e.
the summation of these
 functions divided by ${N_q}$, the total number of such inserted
functions, and let it  have the same dimension in mass and the same order in
$\lambda$ as the original 1PI $n$-point function:
\be
\Gamma^{(n)}(p_1,~\cdots,~p_n; ~q; ~\mu)
=(-i\mu^2)^q(-i\lambda)^{-q}\fr 1 {N_q} \sum
\Gamma^{(n+2q)}(p_1,~\cdots~p_n;~q)
\ee
where $\mu$ is an arbitrary reference mass parameter.  Now we
 evaluate it and analytically continue $q$ to the
complex number.
Then the original potentially divergent 1PI $n$-point function is recovered by
\be
\Gamma^{(n)}(p_1,~\cdots,~p_n;\mu)=\lim_{q\to 0}
\Gamma^{(n)}(p_1,~\cdots,~p_n;~q;~\mu),
\ee
and the original infinity arises manifestly as pole in $q$.

 At the one loop order there are only two divergent graphs, the tadpole
$(t)$ and the fish $(f)$.
In order to regulate the tadpole, we  attach $q$ inserters to
the internal line. Then $(t)$ becomes a $2+2q$-point
function $(t_q)$. For $q$ large enough, $(t_q)$ is
convergent. The regulated function $(t_q')$ is introduced such that it has the
same
 dimension in mass and the same order in $\lambda$ with $(t)$ and
when $q=0$,
$(t_q')\vert_{q=0}=(t)$.
The amplitude of $(t_q')$ can be expressed as {\footnote{In this letter,
the order of the inserted inserters in each inserted graph is always fixed so
that the relevant combinatory factor is  simply fixed to be one as well. One
may relax this simplification and should get similar results.}}
\begin{equation}
(t_q')=(-i\mu^2)^q(-i\lambda)^{-q}(t_q)=\fr{1}{2}\mu^{2q}\int \fr{d^4
l}{(2\pi)^4}
\fr{\lambda}{(l^2-m^2)^{q+1}}.
\end{equation}
It can be easily integrated and expressed
in terms of the $gamma$ functions of $q$:
\begin{equation}
(t_q')=\mu^{2q}\fr{i}{2}\fr \lambda {(4\pi)^2}\fr{\Gamma(q-1)}
{\Gamma(q+1)(-m^2)^{q-1}}.
\end{equation}
Now we analytically continue
  $q$ to the complex number. The original
 tadpole function $(t)$ is then recovered as  the $q\to 0$ limiting case
of $(t_q')$:
\begin{equation}
(t)=\lim_{q\to 0} (t_q')=\fr{i}{2}\fr \lambda {(4\pi)^2}m^2[\fr{1}{q}
+1+\ln (-\fr{\mu^2}{m^2})+o(q)].
\end{equation}

In order to regulate the fish, we attach to its internal lines
$q$ inserters and
it turns to a set of the graphs $(f_{q,i})$ with $i$
inserters inserted in one internal line while $q-i$ in the other.
For $q$ large enough, all $(f_{q,i})$ are convergent.
Then we take their arithmetical average
$(f_q)=\fr 1 {N_q}\sum_{i=0}^q(f_{q,i}), ~{N_q}=q+1$. The regulated
function $(f_q')$ is defined as
\begin{equation}\label{a}\begin{array}{cl}
(f_q')&=(-i\mu^2)^q(-i\lambda)^{-q}(f_q)\\[4mm]
&=\fr{\mu^{2q}}{2(q+1)}\sum_{i=0}^{q}\int \fr{d^4 l}{(2\pi)^4}
\fr{\lambda^{2}}{(l^2-m^2)^{i+1}((p_1+p_2+l)^2-m^2)^{q-i+1}}\\[4mm]
&~~~+(p_2 \to p_3) + (p_2\to p_4).\end{array}\end{equation}
And we can  get
\begin{equation}
\begin{array}{ll}
(f_q')
=\fr{i}{2(4\pi)^2}\lambda^{2}\mu^{2q}\fr{1}{q(q+1)}
\int_0^1d\alpha \fr{1}{[\alpha(1-\alpha)(p_1+p_2)^2-m^2]^q}
+(p_2 \to p_3) + (p_2\to p_4).\end{array}
\end{equation}
We now  analytically continue $q$ to the complex number. Then the fish function
$(f)$ is reached by  the $q\to 0$
limiting case of $(f_q')$,
\begin{equation}
\begin{array}{cl}
(f)&=\lim_{q\to 0}(f_q')\\
&=i\fr {{\lambda}^2}
{(4\pi)^2}[\fr{3}{2q}+\fr{3}{2}+\fr{3}{2}\ln(-\fr{\mu^2}{m^2})+
A(p_1,\cdots, p_4)+o(q)],
\end{array}
\end{equation}
where
$$A(p_1,\cdots, p_4)=
-\fr{1}{2}\sqrt{1-\fr{4m^2}{(p_1+p_2)^2}}
\ln \fr{\sqrt{1-\fr{4m^2}{(p_1+p_2)^2}}+1}{\sqrt{1-\fr{4m^2}{(p_1+p_2)^2}}-1}+
(p_2 \to p_3) + (p_2\to p_4).$$
Thus we complete the regularization of the $\phi^4$ theory at the one loop
order by means of the  inserter method.

Let us now consider how to apply this method to QED. In QED, since the
electron-photon vertex carries a $\gamma$-matrix
and is a Lorentz vector, simply inserting  the vertex would increase the rank
of the functions as  Lorentz tensors and the problem could become quite
complicated. Therefore,
  how to construct a suitable inserter is the first problem to be solved
in addition to the above procedure in the $\phi^4$ theory.
In order to avoid that complication, we borrow an inserter of the Yukawa type
for the massive fermions from the Standard Model
(application of inserter regularization method to Standard Model will appear
elsewhere)
and employ
it for the purpose in QED.
It is an $ff\phi$-vertex of the Yukawa type with a zero
 momentum Higgs external line.  The Feynman rule of such an inserter is
\be
I^{\{f\}}(p)=-i\lambda_f,
\ee
where $\lambda_f$ takes value $\fr g 2  {m_f} /{M_W}$ in the Standard Model,
but
here its value is irrelevant for our purpose.

The divergent 1PI graphs at the one loop order in QED are those contribute
to the
vacuum polarization $\Pi_{\mu \nu}(k)$, the electron self-energy $\Sigma(p)$,
the vertex function $\Lambda_{\mu} (p',p)$ and the photon-photon scattering
function
 which are superficially quadratically, linearly and
logarithmically ultraviolet divergent respectively.\footnote{ For simplicity,
we take the
Feynman gauge $\xi =1$ in this letter.}
Let us now render them finite by means of the inserter procedure.

To regulate the divergent vacuum polarization function $\Pi_{\mu \nu}(k)$,
we attach to one
internal fermion line with $i$ inserters  and to
the other  with $2q-i$ ones. Then we get
a set of $2+2q$-point functions $\Pi^{\{q,i\}}_{\mu \nu}(k; q)$. If $q$ is
large enough, all these
$2+2q$-point functions are convergent. Then we introduce a new function
\be
\Pi_{\mu \nu}(k;q;\mu)=(-i\mu)^{2q} (-i\lambda_f)^{-2q} \fr 1 {N_q}
\sum_{i=0}^{2q}
\Pi^{\{q,i\}}_{\mu \nu }(k; q),
\ee
which has the same dimension in mass, the same order in $e$ with the original
function $\Pi_{\mu
\nu}(k)$. It is not hard to prove that this  function can be  expressed as
$$
\Pi_{\mu \nu}(k;q;\mu)=-\mu^{2q}e^{2}\fr 1 {N_q} \sum_{i=0}^{2q}\int \fr{d^4
p}{(2\pi)^4}
Tr[\gamma_{\mu}\Bigl(\fr{1}{\not{p}-\not{k}-m}\Bigr)^{i+1}
\gamma_{\nu}\Bigl(\fr{1}{ \not {p} -m}\Bigr)^{2q-i+1}]
$$
and satisfies the gauge invariant condition:
\be
k^{\mu}\Pi_{\mu \nu}(k;q;\mu)=0.
\ee
Continuing $q$ to the complex number, thus the original amplitude
$\Pi_{\mu\nu}$ is recovered as
\be
\Pi_{\mu\nu}(k)=\lim_{q\to 0}\Pi_{\mu\nu}(k;q;\mu).
\ee
If we denote
\be
\Pi_{\mu\nu}(k^2)\equiv
(k_{\mu}k_{\nu}-k^2g_{\mu\nu})\Pi(k^2),~~~\Pi(k^2)=\Pi(0)+\Pi^f(k^2),
\ee
by some calculation, we get
\be\begin{array}{ll}
\Pi(0)
=e^2\fr{4i}{(4\pi)^2}[\fr{1}{3q}
+C+\frac{1}{3}\ln(-\fr{\mu^2}{m^2}) +
o(q)],\\[4mm]
\Pi^f(k^2)=-\fr{ie^2}{2\pi^2}
\int_0^1 d\alpha \alpha(1-\alpha)\ln[1-\fr{\alpha(1-\alpha)k^2}{m^2}].
\end{array}\ee
where $C$ is some constant. The finite part $\Pi^f(k^2)$ is the same as that
derived in other regularization procedures.

To regulate the electron self-energy function $\Sigma(p)$, we attach to the
internal fermion line with $2q$ inserters and the
graph $\Sigma(p)$ is turned to a $2+2q$-point convergent function
 $\Sigma(p; q)$ if $q$ is large enough. Then we introduce a
new function:
\be
\Sigma(p; q;\mu)=(-i\mu)^{2q} (-i\lambda_f)^{-2q}\Sigma (p;q),
\ee
which can be expressed as
$$
\Sigma(p; q;\mu)_{\beta\alpha}=-\mu^{2q}e^2
\int \fr{d^4 k}{(2\pi)^4}
\Bigl(\gamma^{\mu}\fr{(\not{k}+m)^{1+2q}}{(k^2-m^2)^{1+2q}}
\gamma_{\mu}\Bigr)_{\beta\alpha}\fr{1}{(p-k)^2}.
$$
Continuing $q$ to the complex number, the original function  $\Sigma(p)$ is
 reached by
\be
\Sigma(p)_{\beta\alpha}=\lim_{q\to 0}\Sigma(p; q;\mu)_{\beta\alpha}.
\ee
Denoting
\be
\Sigma(p)=mA(p^2)+iB(p^2)\not{p},
\ee
we may finally get
$$
\begin{array}{cl}
A(p^2)=&-\fr{ie^2}{4\pi^2}\{\fr{1}{q}+3-\ln\fr{p^2-m^2}{\mu^2}
+\fr{m^2}{p^2}\ln(1-\fr{p^2}{m^2})+A^f(p^2)\},\\[4mm]
B(p^2)=&\fr{e^2}{(4\pi)^2}\{\fr{1}{q}+\fr{1}{2}+\fr{m^2}{p^2}
-\ln\fr{p^2-m^2}{\mu^2}
-(\fr{m^2}{p^2})^2\ln(1-\fr{p^2}{m^2})\}+B^f(p^2),
\end{array}
$$
both $A^f(p^2)$ and $B^f(p^2)$ are finite function.

Similarly,  to regulate the  vertex function,
 $\Lambda_\mu (p', p)$, we attach to the internal fermion lines with $2q$
inserters to  get a set of $(3+2q)$-point functions
$\Lambda^{\{q,i\}}_\mu(p', p; q)$. Then we introduce
\be
\Lambda_\mu(p', p; q;\mu)=(-i\mu)^{2q} (-i\lambda_f)^{-2q}\fr 1 {N_q}
\sum_{i=0}^{2q}
\Lambda^{\{q, i\}}_{\mu}(p',p; q),
\ee
which can be expressed as
$$
\Lambda_\mu(p', p; q;\mu)=-\fr {\mu^{2q}e^{3}} {2q+1} \sum_{i=0}^{2q}
\int \fr{d^4 l}{(2\pi)^4}
\Bigl(\gamma^{\rho}\fr{(\not{l}-\not{k}+m)^{1+i}}
{[(l-k)^2-m^2]^{i+1}}\gamma_{\mu}
\fr{(\not{l}+m)^{1+2q-i}}{(l^2-m^2)^{1+2q-i}}\gamma_{\rho}
\Bigr)
\fr{1}{(p-l)^2}.
$$
Continuing $q$ to the complex number,  the original vertex function is then
recovered
as:
\be
\Lambda_\mu(p', p)=\lim_{q\to 0}\Lambda_\mu(p', p; q;\mu).
\ee
Finally, we find that
\be\begin{array}{ll}
\Lambda_\mu(p', p)
=\fr{-ie^3\mu^{2q}}{(4\pi)^2}\gamma_{\mu}~~~~~~~~~\\[4mm]
{}~~[\fr{1}{q}-\fr{3}{2}-
\int_0^1 d\alpha\int_0^{1-\alpha}d\beta
\ln \{\beta(1-\beta)k^2+\alpha(1-
\alpha)p^2-2\alpha\beta k\cdot p-(1-\alpha)m^2\}]\\[4mm]
{}~~-\fr{ie^3}{2(4\pi)^2}\int_0^1 d\alpha\int_0^{1-\alpha}d\beta \fr
{\gamma^{\rho}[(\beta-1)\not{k}+\alpha \not{p}+m]\gamma_{\mu}
[\beta\not{k}+\alpha\not{p}+m]\gamma_{\rho}}{\beta(1-\beta)k^2+\alpha(1-
\alpha)p^2-2\alpha\beta\cdot p-(1-\alpha)m^2}+o(q).
\end{array}
\ee
The observable part $\Lambda_{\mu}^f$  of the vertex function is
defined by
$$
\Lambda_{\mu}=K\gamma_{\mu}+\Lambda_{\mu}^f,
$$
where $K$ contains the pole in $q$ when $q\to 0$ and $\Lambda_{\mu}^f$ is
finite from which the anomalous magnetic moment of the
electron can be derived. The result is the same as in other approaches.

Similar procedure may also be applied to
the photon-photon scattering
$\Gamma_{\mu \nu \rho \sigma}(p_1 \cdots p_4)$. It is finite due to the gauge
invariance.
We now check its gauge invariance
by the inserter method.
Attaching $2q$ inserters to internal fermion lines in all possible ways
and introducing the regulated function
\be
\Gamma_{\mu \nu \rho \sigma}(p_1 \cdots p_4;q;\mu)
=(-i\mu)^{2q} (-i\lambda_f)^{-2q}
\fr 1 {N_q}\sum_{i=0}^{2q}\sum_{j=0}^{2q-i}
\sum_{l=0}^{2q-i-j} \Gamma^{\{q,i,j,l\}}_{\mu \nu \rho \sigma }
(p_1 \cdots p_4),
\ee
 it can be proved that this function satisfies the gauge invariant condition.
Continuing
 $q$ to the complex number, the
 original  function is then recovered by
\be
\Gamma_{\mu \nu \rho \sigma}(p_1 \cdots p_4)
=\lim_{q\to 0}\Gamma_{\mu \nu \rho \sigma}(p_1 \cdots p_4; q;\mu).
\ee
By some straightforward calculation, we may explicitly show that
\be
\Gamma_{\mu \nu \rho \sigma}(p_1 \cdots p_4)\vert_{p_1= \cdots =p_4=0}=0.
\ee
This also coincides with the gauge invariance. Thus the inserter regularization
does preserve the gauge symmetry in QED at the one loop level.

We now apply the inserter method to some
SUSY-models at one-loop order. We will show that the SUSY-version of the
inserter method does preserve
supersymmetry manifestly and consistently by reexamining some well-known and
simple examples at
the one loop order.

Let us first consider an example in the massive Wess-Zumino model.
At the one
loop level, the self-energy graph of antichiral-chiral superfield propagator
$\bar\phi \phi$
 is divergent. After some $D$-algebraic manipulation, it is left a divergent
 integral
\be
\int d^4\theta ~\phi(-p, \theta) \bar \phi(p, \theta)~A(p, m),
\ee
where
\be
A(p, m)
=\int \fr{d^4 k}{(2\pi)^4}
\fr{1}{(k^2+m^2)((p+k)^2+m^2)}.
\ee
To regulate this integral, we need first to
 construct an antichiral-chiral superfield inserter. For such an inserter, we
take a pair of vertices linked
by a $\bar\phi \phi$-internal line with a pair of chiral and antichiral
external
legs carrying zero momenta. Its Feynman rule can easily be written down.
Now we may utilize this inserter to insert $q$-times the internal lines in the
divergent graph. Then we get a set of convergent graphs with $i$-inserters in
one internal line and $q-i$ in the other. Similarly, after  some $D$-algebraic
manipulation, the corresponding convergent function $I^{\{q,i\}}(p)$ is
proportional to
\be
\int d^4\theta~\phi(-p, \theta)  \bar \phi(p, \theta)~A^{\{q,i\}}(p, m),
\ee
where
\be
A^{\{q,i\}}(p, m) =\int \fr{d^4 k}{(2\pi)^4}
\fr{1}{(k^2+m^2)^{2i+1}((p+k)^2+m^2)^{2q-2i+1}}.
\ee
It is very similar to the ones in the case of inserted fish functions in the
$\phi^4$ theory(see (\ref{a})), so we will not repeat it
 here.

Let us consider a most general
$N=1$ supersymmetric  renormalizable model which is invariant under a gauge
group $G$
contains chiral superfields $\phi^a$ in a representation $R$ of $G$ and vector
superfield $V$.
The one-loop correction to the $\phi^a \bar \phi_a$ propagator are given by two
divergent graphs. One is the same as in the Wess-Zumino model while the other
is the one with an internal antichiral-chiral line replaced by a $V$-line.
They lead to the expression:
\be
g^2 \int d^4\theta ~\phi^b(-p, \theta)
(S^a_b-C_2(R) \delta^a_b) \bar \phi_a(p, \theta)~A(p),
\ee
where $A(p)=A(p, m=0)$ and
$$ 2g^2 S^a_b=d^{ace}d_{bce},$$
$d^{ace}$ are the couplings of  the $\phi^3$-term. The cancellation condition
$S^a_b=C_2(R) \delta^a_b$ should hold much safer if the divergent integrals
 $A(p)$
in two graphs can be regulated in a way of preserving supersymmetry manifestly
and consistently. This can be done by means of the SUSY-version of the inserter
method.
To this end, in addition to the inserter for the chiral superfield constructed
above ( the internal representation indices should be paired here ),
we need an inserter for the $V$-internal line as well. In fact,
it may be constructed in such a way
that two $V\phi \phi$-vertices linked by an internal antichiral-chiral line
with two external
$\phi^a$, $\bar \phi_b $ legs carrying zero momenta and paired representation
indices. Then it is easy to see that by inserting these two inserters in the
internal $V$ line and the internal
  antichiral-chiral line respectively, we can always get the same regulated
functions for the both graphs. Therefore, the cancellation can be insured at
the regulated function level as well.

 The one-loop correction to the vector superfield propagator is given by three
 graphs with $V$-loop, $\phi$-loop and the ghost-loop respectively. The
SUSY-version of the inserter method also ensures the corresponding cancellation
condition holds at the regulated function level as long as we employ the ghost
inserter as a pair of the  ghost-V vertices linked by an internal ghost
antichiral-chiral  line with
two external $V$ legs carrying zero momenta in addition to the fore-mentioned
inserters for the internal $V$ line and the internal antichiral-chiral line.

The SUSY-version of the inserter method may also be combined with the
background approach. For example, in the background field approach
the above one-loop contribution to the $V$ self-energy from a massive chiral
superfield
leads to one divergent integral only:
\be
\fr 1 4 C_2 ({R})~ tr
\int d^4\theta ~W^\alpha (p, \theta ) \Gamma_\alpha( -p, \theta)~A(p, m),
\ee
where $W^\alpha (p, \theta )$ is the superfield strength and $ \Gamma_\alpha(
-p, \theta)$ the background field connection. Again, we may utilize the
an antichiral-chiral superfield inserter
 in the background field to get a set
of convergent integrals and to regulate this divergent one in the way of
preserving supersymmetry manifestly and consistently.

Finally, let us make some remarks for the inserter regularizaton method.

 The crucial point of
this approach is very simple but fundamental. That is, the entire procedure
 is intrinsic in the QFT. There is nothing
changed, the action, the Feynman rules, the spacetime dimensions etc.
are all the same as that in the given QFT. Although for QED, the inserter we
have employed is borrowed from the Standard Model, QED is in fact unified with
the weak interaction in the Standard Model. Therefore, it is still intrinsic
in the Standard Model. Consequently, in applying to other cases all
symmetries and
topological properties there should be preserved  in principle, especially to
those cases
where the symmetries and topological properties are sensitive to the
spacetime dimensions,
the number of
 fermionic degrees of freedom.
Of course, for each case some special care should be taken.
 We have shown that for some SUSY-models at one loop order it is the case.
We will study  higher loop orders and other cases in detail elsewhere.

In our method the inserters play an important role. Of course, the
zero-momentum-leg(s)
in the inserters do not represent realistic particles. But, the method  may
have some physical
explanation. Namely, for each inserted internal line, the virtual particle
 always
emits
 and/or absorbs  via the
 inserters other far-infrared ``particles'' that carry zero momenta from
the vacuum. In other words, the vacuum
is full of such far-infrared ``particles'' that they always have or pair
 together with the vacuum quantum numbers, i.e.  zero momenta,  singlet(s) in
all internal symmetries ( including gauge
symmetries ) and scalar(s) in the spacetime symmetries.
The  inserter method  indicates that the pathological behavior of
 those divergent
graphs can be remedied by taking into account the role played by these
 far-infrared
``particles''.

 We have not devoted any attention in this letter to the infrared divergences
at all. This issue is in fact another most challenging problem to all known
regularization
schemes.  It is intriguing to see, however, as far as the vacuum picture
 is concerned,  certain kind of inserters should be constructed and some
intrinsic relation between
the divergent function and convergent ones may also be established in the
 infrared region. Then
the intrinsic inserter method may work in this region as well. We will
also
investigate this issue elsewhere.

\vskip 7mm
{\it  The work by HYG was mostly done during his visiting to The
Max-Planck-Institut
f\"ur Mathematik, Bonn. He would like to thank Professors F. Hirzebruch  for
warm hospitality. He is also grateful to Professor
W. Nahm for valuable discussion and warm hospitality.
 This work is supported
in part
by The National Natural Science Foundation of China.}

\vskip 8mm

\bigskip\bigskip
{\raggedright \bf References\\}
\begin{description}

\item[1] Zhong-Hua Wang  and Han-Ying Guo, Intrinsic vertex regularization and
renormalization in $\phi^4$ theory. 1992. ITP-CAS and SISSA preprint.
 Unpublished.

\item[2] Zhong-Hua Wang  and Han-Ying Guo, Intrinsic loop regularization and
renormalization in $\phi^4$ theory. Comm. Theor. Phys. ( Beijing ) {\bf 21}
(1994) 361.

\item[3] Zhong-Hua Wang  and Han-Ying Guo, Intrinsic loop regularization and
renormalization in QED. To appear in Comm. Theor. Phys.  ( Beijing ).

\item[4] Zhong-Hua Wang  and Luc Vinet, Triangle anomaly from the point of view
of loop regularization. 1992. Univ. de Montr\`eal preprint. Unpublished.

\item[5] Dao-Neng Gao, Mu-Lin Yan and Han-Ying Guo,  Intrinsic loop
regularization in quantum field theory. To appear in the Proc. of ITP Workshop
 on QFT (1994).

\end{description}

\end{document}